\documentclass[prd, showpacs, 
               nofootinbib, preprintnumbers, floatfix]{revtex4}%
\usepackage{epsfig}
\usepackage{amsmath}

\begin{document}
\newcommand{\ignore}[1]{}

\unitlength=1.2mm \preprint{NT@UW-11-15}

\title{Roper Properties on the Lattice: An Update}


\author{Huey-Wen Lin}
\email{hwlin@phys.washington.edu}
\affiliation{Department of Physics, University of Washington, Seattle, WA 98195}

\author{Saul D. Cohen}
\affiliation{Department of Physics, University of Washington, Seattle, WA 98195\\
Center for Computational Science, Boston University, Boston, MA 02123
}

\begin{abstract}
In this proceeding, I give an overview of the latest progress in lattice-QCD calculations of the Roper resonance, and Roper-nucleon electromagnetic and axial transition form factors using dynamical gauge ensembles.
\end{abstract}

\maketitle

{\bf Introduction: }%
The nature of the Roper resonance, $N(1440)\ P_{11}$, has been the subject of interest since its discovery in the 1960s. The mass of this positive-parity excited state is lower than the negative-parity ground state $N(1535)\ S_{11}$, which is quite unusual, never being seen in mesons. This has led phenomenologists to postulate that the Roper might not simply be a radial excitation of the nucleon. Rather, it might be a hybrid state that couples predominantly to currents with some gluonic contribution 
or a five-quark (meson-baryon) state. 
Experiments at Jefferson Laboratory, MIT-Bates, LEGS, Mainz, Bonn, GRAAL, and Spring-8 offer new opportunities to understand in detail how nucleon resonance properties emerge from nonperturbative aspects of quantum chromodynamics (QCD).

On the theoretical side, many attempts try not only to reproduce the reversal of the mass ordering of Roper and $S_{11}$ starting from basic QCD properties, but also to determine the Roper-nucleon transition form factors. Among these theoretical efforts, lattice QCD (LQCD) is the most promising candidate for studying the nonperturbative aspects of QCD, where the Roper mystery lies. 
Lattice QCD has been successfully employed to study various strong-interaction phenomena, such as the spectroscopy of heavy-quark hadrons, the tower of excited baryon states, flavor physics involving the CKM matrix, hadron decay constants and the strong coupling constant.
In many cases, it can provide higher-precision data from the Standard Model than what can be measured experimentally. 
By discretizing spacetime into a four-dimensional grid and working in Euclidean space, we are able to compute the path integral (in terms of discretized versions of the QCD Lagrangian and operators) directly via numerical integration, providing first-principles calculations of the consequences of QCD, without making uncontrollable approximations. 
The four-dimensional lattice size, lattice spacing, the quark masses in the sea sector and precision of the integration have been greatly improved over the past decade due to increasing computational resources and more efficient algorithms. We can systematically study the effects due to having a finite volume and nonzero lattice-spacing and remove them by extrapolation, using calculations with differing values of these quantities.

Due to limited computational resources, many existing calculations have had to use multiple unphysical pion masses and extrapolate to the physical pion mass via chiral effective theory. (This helps to determine the low-energy constants in the theory.)
The uncertainties caused by the extrapolation have been greatly reduced as lighter pion masses are included in calculations. 
Recently, major collaborations, such as BMW, PACS-CS and MILC have begun generating QCD gauge ensembles at the physical pion mass, which will remove such systematics and put stringent tests on low-energy effective theories.
Lattice calculations with sub-percent precision have been achieved for many flavor-related quantities and used to guide tests of the Standard Model and to probe possible new physics. Calculations of hadron structure and interactions require more computational resources but should soon enter an era of precision.

\begin{figure}
\includegraphics[height=.22\textheight]{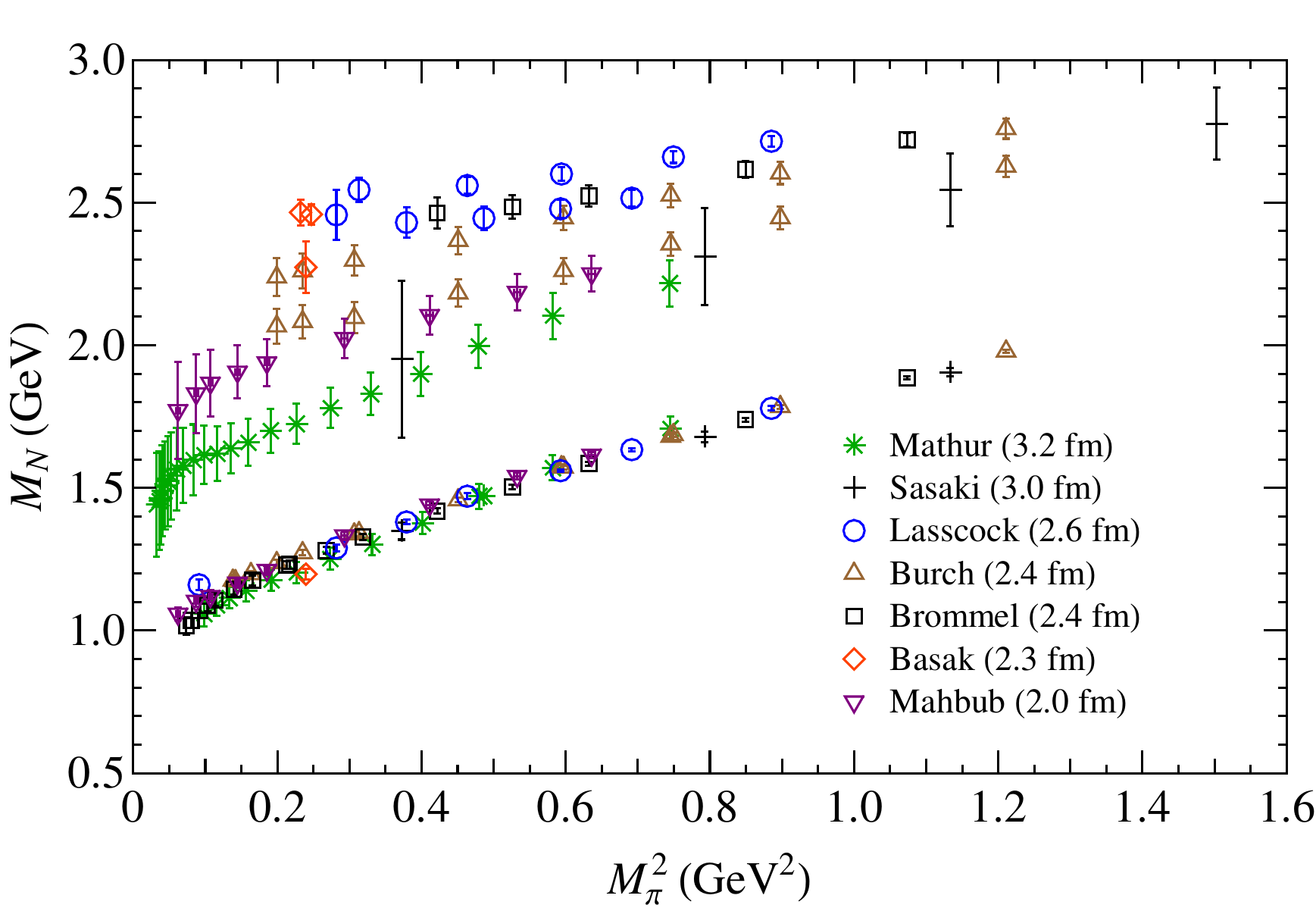}
\includegraphics[height=.22\textheight]{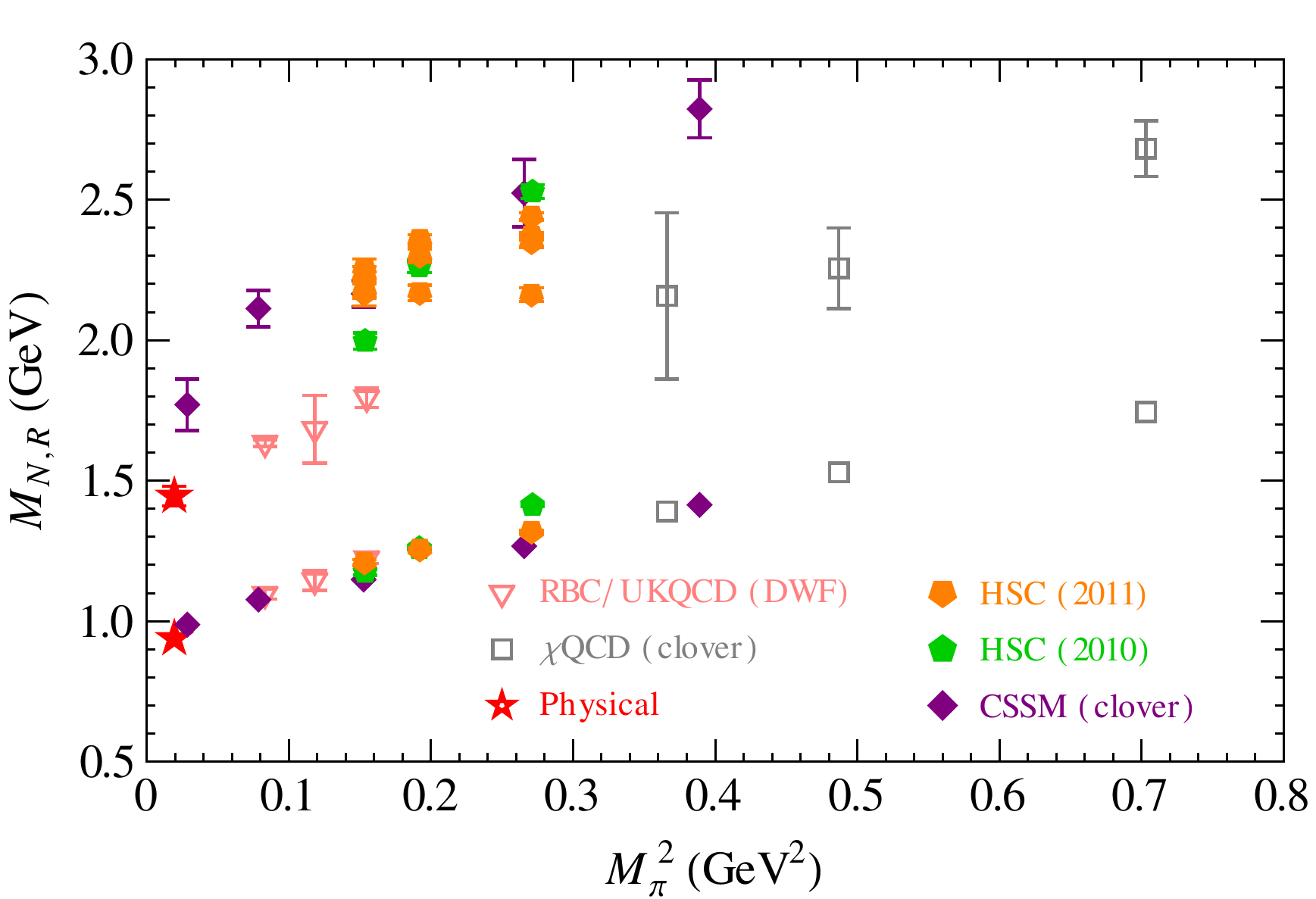}
\caption{\label{fig:ropers}
(left) Summary of published $N_f=0$ LQCD calculations of the nucleon and Roper masses in GeV as functions of $M_\pi^2$.  Note that the errorbars are only statistical.
(right) Summary of the latest $N_f=2+1$ Roper results as a function of pion mass squared.
Note that the filled symbols and solid errorbars (open symbols and dashed errorbars) indicate the results taken from published papers (the latest Lattice Conference proceedings).
}
\end{figure}

{\bf Resonance: }%
Much progress has been made on the Roper resonance.
The left-hand side of Fig.~\ref{fig:ropers} was a state-of-the-art summary of lattice calculations of the Roper resonance as of the last Nstar workshop. Many groups have developed their analyses using the cheapest available lattices (having sufficiently large volume, fine lattice spacing, etc).
However, most of these calculations have been done in the ``quenched'' approximation, where the effective sea-quark mass is infinite; that is, only gluonic interactions with the valence quarks are taken into account. Such an approximation preserves certain QCD properties, such as confinement, asymptotic freedom and spontaneous chiral symmetry breaking, making quenched simulations useful for testing new lattice techniques. However, quenching introduces uncontrollable sources of systematic error that are expected to be 30\% or larger, depending on the physical quantity in question.

In recent years, newer calculations have not needed to apply this approximation, and more results with $N_f=2+1$ (degenerate up and down plus strange in the sea) have become more widely available.

Two preliminary studies were reported in the annual Lattice Conference using $N_f=2+1$ domain-wall fermion and clover lattices by RBC/UKQCD and $\chi$QCD; their results are shown as the open symbols in Fig.~\ref{fig:ropers}, but no refined numbers have been reported since.
The CSSM group\cite{Mahbub:2010rm} used local baryon operators with different smearing parameters to construct a large basis for the nucleon with spin 1/2 and analyzed the data using the variational method. They use $N_f=2+1$ isotropic lattices at a fixed lattice spacing, $a=0.0907$~fm and fixed volume, $32^3\times 64$ with lightest pion mass 156~MeV. Their calculation of the Roper mass is shown on the right-hand side of Fig.~\ref{fig:ropers} as 5 purple diamond points. However, the lightest two pion masses, since their $M_\pi L$ values are less than 3, may suffer finite-volume effects (which have not currently been estimated).
Hadron Spectrum Collaboration (HSC) reported two studies using a large basis of independent operators organized according to representations of the cubic group, including operators where the three quarks inside the nucleon are located at different lattice sites\cite{Bulava:2010yg,Edwards:2011jj}. They have used $2+1$-flavor anisotropic clover lattices, with finer temporal lattice spacing to improve the excited-state signal in Euclidean time; that is, the lattices have better resolution in the direction from which the particle energies are extracted. The green pentagons in right-hand side of Fig.~\ref{fig:ropers} are taken from the lowest two states reported in their study last year using only operators in the $G_{1g}$ irrep. This year, a refined study mapped the cubic-irrep operators to definite spins using derivative operators constructed with continuum symmetries. They found four states closely aligned and roughly around the excited state found by CSSM.
A refined CSSM study reported in this year's Nstar Workshop\footnote{See proceeding by D.~Leinweber.} using a larger basis of smeared operators that they found a similar structure as Ref.~\cite{Edwards:2011jj}.
This may suggest that multiple-particle operators need to be included in the correlator matrix to clearly identify the nature of these closely related states; both HSC and CSSM are currently investigating this.

The dynamical results in Fig.~\ref{fig:ropers} are impressive, but the errorbars shown above are only statistical; that is, there is no estimation of the systematics due to the finite volume and nonzero lattice spacing in the calculations. This is sometimes difficult to do due to the limitations of the dynamical lattices available. 
Ref.~\cite{Beane:2011pc} has demonstrated with high-statistics calculations using $N_f=2+1$ anisotropic clover lattices at $M_\pi \approx 390$ MeV. Four volumes are used in this calculation: 2, 2.5, 3 and 4~fm with $M_\pi L \approx 3.9$, 4.8, 5.8, 7.7. The volume dependence is not noticeable in the earlier HSC study\cite{Lin:2008pr} with less trajectories available. To our surprise, the nucleon ground state with $M_\pi L$ around 4, which is usually the rule of thumb for where finite-volume effects can be ignored, shows around a 30-MeV discrepancy from its value in the infinite-volume limit.

We investigate the effects of lattice volume on the Roper mass and found the effects to be even more dramatic; we observed the central-value shift to be about 170 MeV. 
More time should be devoted to studying volume dependence to get a better estimation of the finite-volume effects in these dynamical calculations.  Furthermore, we found the overlap factors associated with the Roper states at the largest two volumes, gave a ratio $A_R(4~{\rm fm})/A_R(3~{\rm fm})=0.964(52)$, consistent with one; this suggests that the first-excited state is likely to be a one-particle state rather than a scattering state for the ensembles with pion mass larger than 390~MeV.

{\bf Electromagnetic Transition Form Factors: }%
Roper-$N$ transition form factors provide a different approach to understanding the unusual nature of the Roper state. They also provide information about the spatial distribution of the hadron. The general spin-1/2 electromagnetic form factors can be obtained from the following matrix elements:
\begin{eqnarray}
\label{eq:Vector-roper}
\langle N_2\left|V^{\rm }_\mu\right|N_1\rangle(q) &=&
{\overline u}_{N_2}(p^\prime)\left[ F_1^*(q^2) \left(\gamma_{\mu}-\frac{q_\mu}{q^2}
q\!\!\!/
\right)
+ \sigma_{\mu \nu}q_{\nu}
\frac{F_2^*(q^2)}{M_{N_1}+M_{N_2}}\right]u_{N_1}(p),
\end{eqnarray}
where $q=p^\prime-p$.
On the lattice, we can obtain both the form factors related to the Roper decay $P_{11} \rightarrow \gamma N$, and the one that is related to photoproduction, $\gamma^* N \rightarrow P_{11}$.

We performed the first LQCD calculation of the Roper-nucleon transition form factors a few years ago, also starting with the quenched approximation where the QCD vacuum only contains gluons, using single pion mass around 720~MeV; see Figs. 1, 2 in Ref.~\cite{Lin:2008qv}. 
Since the nucleon and Roper masses are much higher than the physical ones, this calculation is in the time-like region for the matrix element $\langle P_{11} |V_\mu | N \rangle$. When we decrease the pion mass, we will enter the space-like region and this matrix element will be helpful in giving us different $Q^2$ points. 
We found there are significant differences from the experimental values in the low-$Q^2$ region, which might be caused by the heavier pion mass used in the calculation or by the quenched approximation.

A later study explored the pion-mass dependence of the transition form factors $F_{1,2}^*$ at three pion masses, as well as large-$Q^2$ form factors; see Fig.~4 in Ref.~\cite{Lin:2008gv}.
In the time-like $Q^2$ regions, all four transition form factors have very mild quark-mass dependence. Even in the space-like region, the dependence only becomes evident at $Q^2>2$~GeV$^2$.
We obtain form factors with momentum transfer up to 6~GeV$^2$; 
in the large-$Q^2$ region, our calculations show that these form factors have about the same magnitude as the single-pion or two-pion analyses of the experimental data. However, in the low-$Q^2$ region, both the CLAS analysis and PDG value (at zero momentum transfer) still have opposite sign from ours for both proton and neutron transition form factors.

\begin{figure}
\includegraphics[height=.22\textheight]{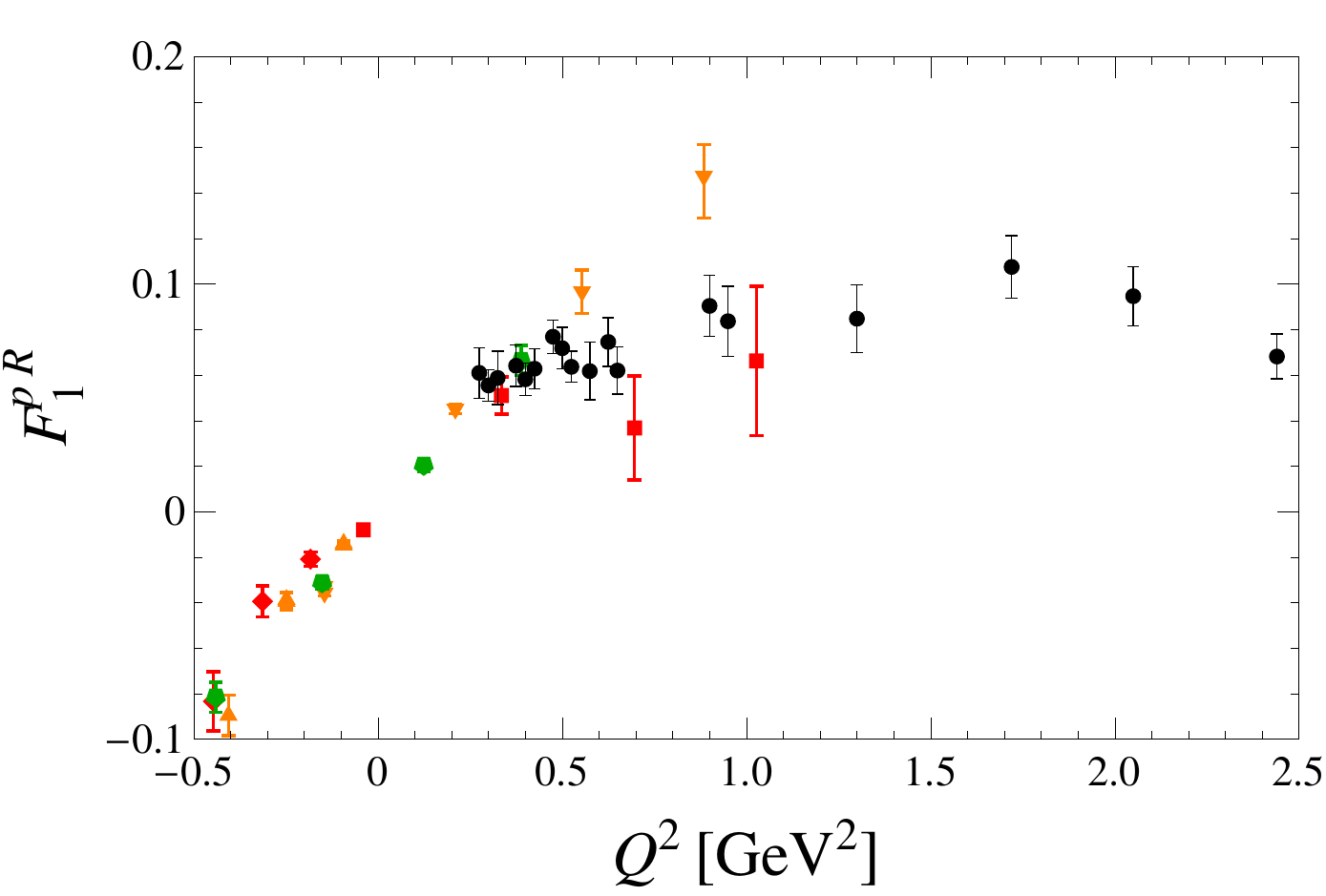}
\includegraphics[height=.22\textheight]{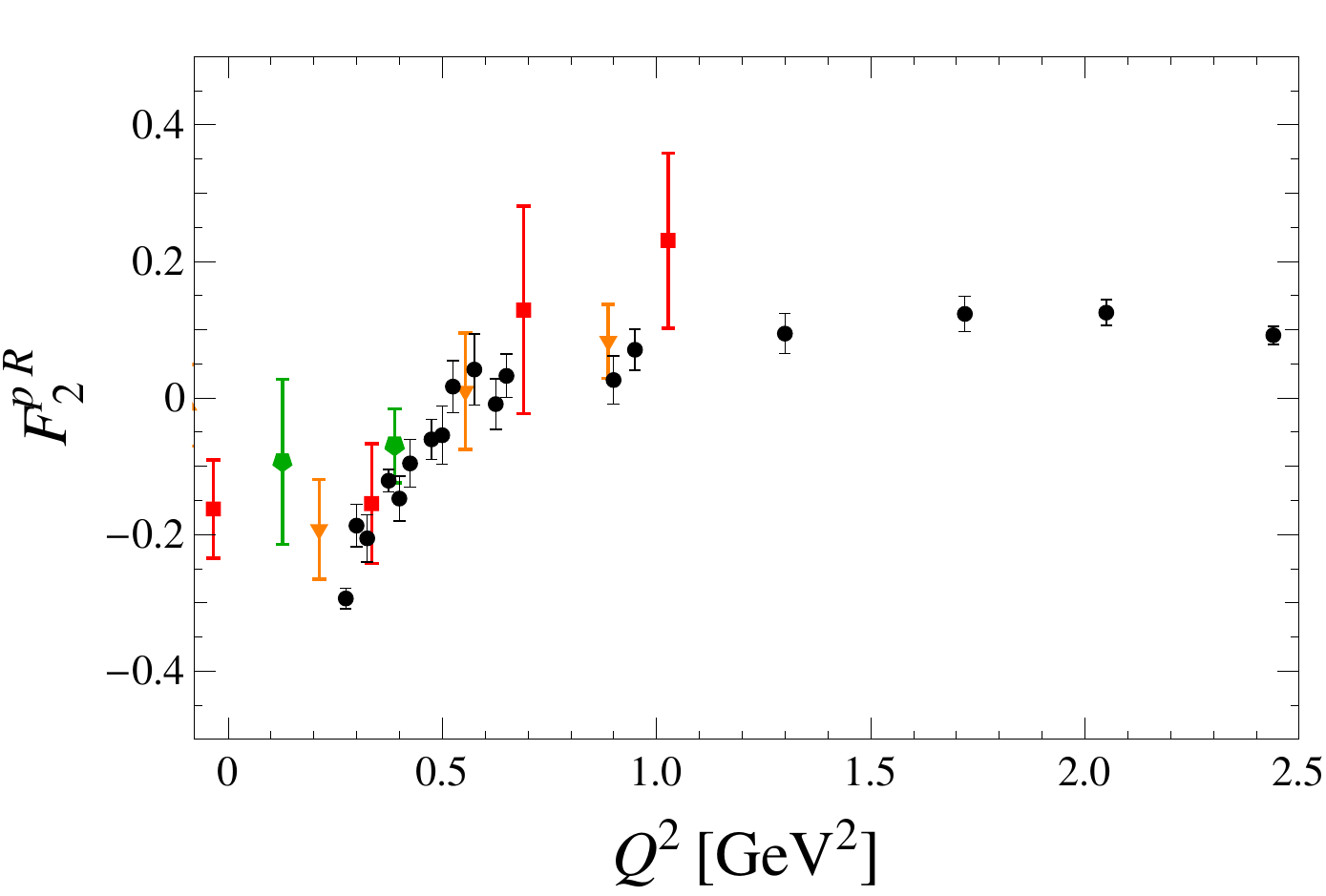}
\caption{\label{fig:F12pR-dyn}
Proton-Roper transition form factor $F_1^{pR}(Q^2)$ (left) and $F_2^{pR}(Q^2)$ (right) on the $N_f=2+1$ anisotropic lattices with $M_\pi\approx$ 390, 450, 875~MeV whose volumes are 3, 2.5, 2.5~fm, respectively.
}
\end{figure}

In this workshop, we report preliminary results using $N_f=2+1$ anisotropic dynamical lattice, focusing on the low-$Q^2$ region. 
Figure~\ref{fig:F12pR-dyn} shows three pion masses at a volume of 3, 2.5, 2.5~fm, where the volume dependence is not significant for Roper states; the experimental values are reconstructed using experimental helicity amplitudes
from CLAS~\cite{Mokeev:2011,Mokeev:2010va,Aznauryan:2009mx,Dugger:2009pn,Mokeev:2008iw}.
At the current statistics, the data indicate a sign change in the form factor relative to the quenched approximation, suggesting that in the low-$Q^2$ region, the pion cloud effects are significantly larger. Thus, removal of the quenched approximation causes a sign change.

{\bf Axial Transition Couplings: }%
The axial transition form factors contain spin-structure information about hadrons, and since the Roper is the lowest excited state of the nucleon, it is important to obtain this information for use in experimental analysis. Processes such as neutrino-nucleus scattering that are used to extract neutrino mass splittings and mixing angles require such inputs to control systematic errors.

\begin{figure}
\includegraphics[height=.22\textheight]{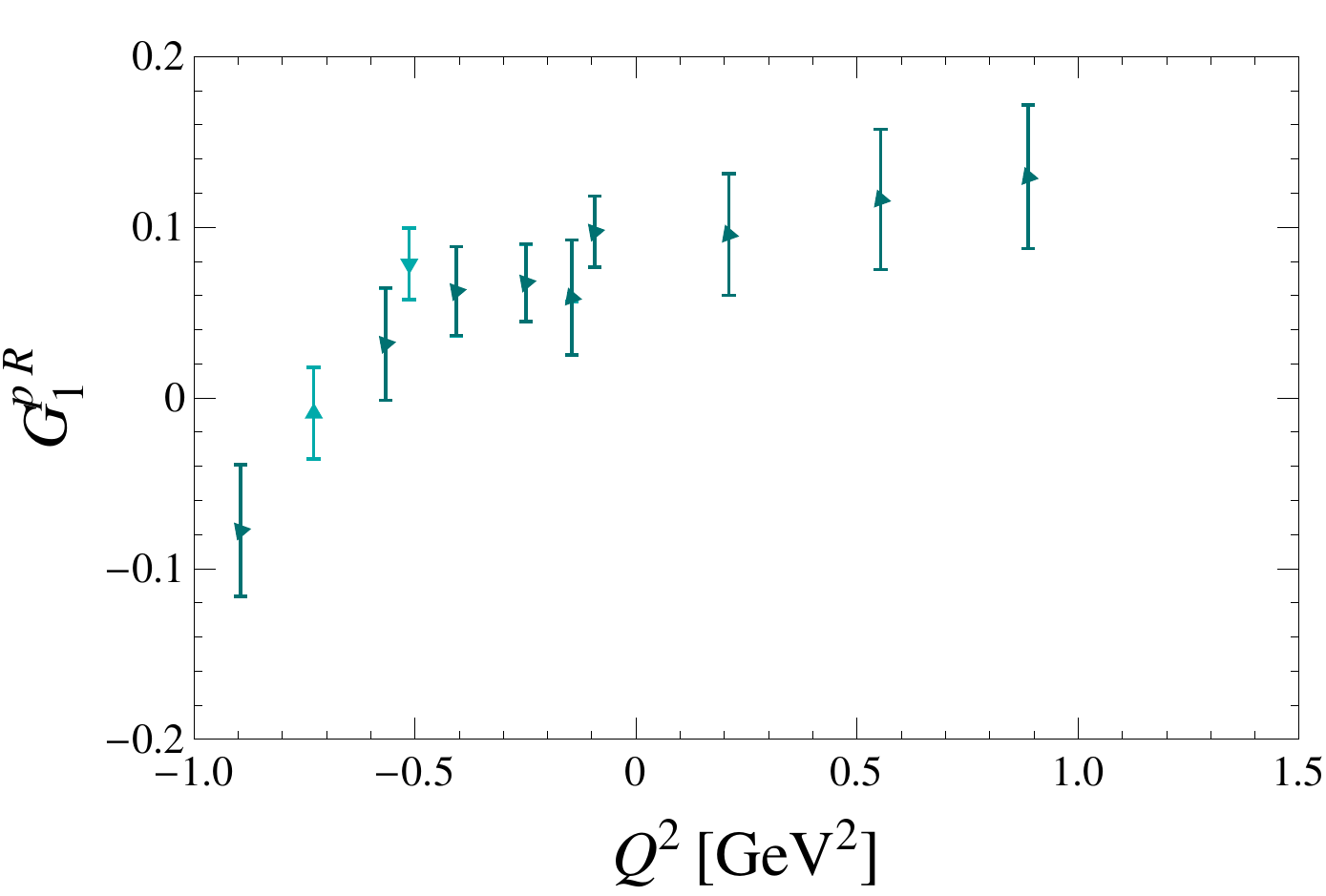}
\includegraphics[height=.22\textheight]{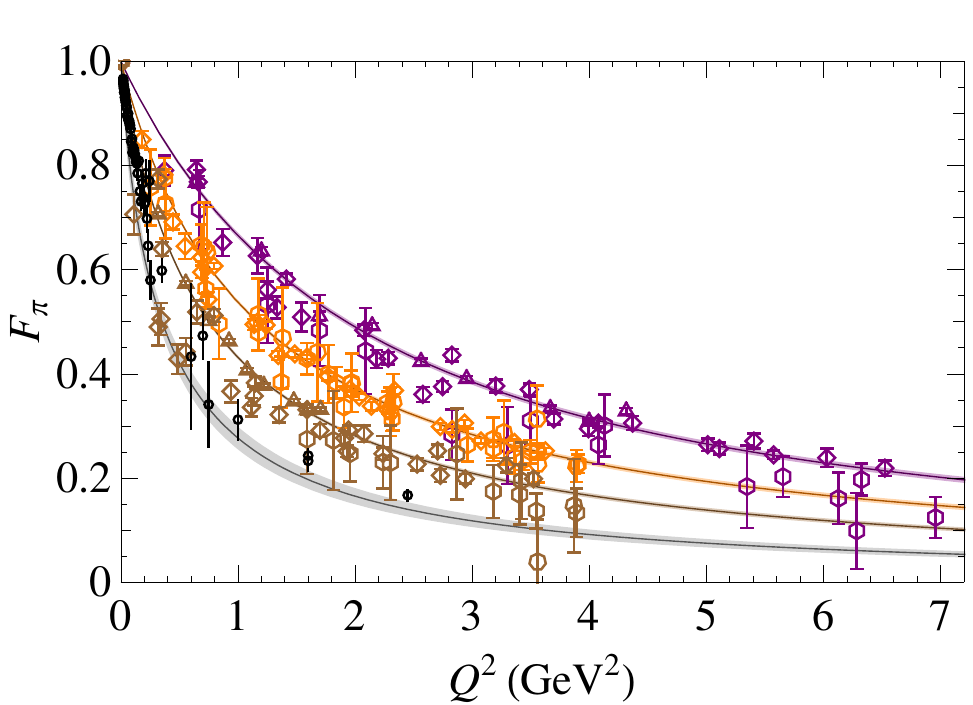}
\caption{\label{fig:G1pR-Fpi}
(Left) Axial transition form factor $G_1^{pR}(Q^2)$ on the $N_f=2+1$ anisotropic lattices with $M_\pi\approx$ 390, 450, 875~MeV.
(Right) Pion form factor with multiple pion masses at 580, 875, 1350~MeV. The experimental points are shown as (black) circles while the lowest gray band is the extrapolation to the physical pion mass using our lattice points.
}
\end{figure}

There are two approaches to extracting axial transition couplings on the lattice.
A direct calculation of the transition form factor through the matrix elements
\begin{eqnarray}
\label{eq:Axial-roper}
\langle N_2\left|A^{\rm }_\mu\right|N_1\rangle(q) &=&
{\overline u}_{N_2}(p^\prime)\left[ G_1(q^2) \gamma_{\mu}\gamma_{5}
+ q_{\mu}\gamma_5
\frac{G_2(q^2)}{M_{N_1}+M_{N_2}}
+\gamma_5 (p+p^\prime)_\mu \frac{G_3(q^2)}{M_{N_1}+M_{N_2}}
\right]u_{N_1}(p)
\end{eqnarray}
at various momenta allows extraction of $G_1(q^2)$ and extrapolation to zero transfer momentum. We carry out the first lattice calculation of the axial proton-Roper transition form factor on the same 2+1-flavor anisotropic lattices at the heaviest pion mass of 875~MeV; see the left-hand side of Fig.~\ref{fig:G1pR-Fpi}. Our current estimation of $G_1(0)$ is of the order of 0.1.

Alternatively, we can study the volume-dependence of the nucleon mass, and use finite-volume chiral perturbation theory to extract the coupling constants.
For example, NPLQCD studied the volume dependence of octet baryons at high precision.
The axial decuplet--octet-meson couplings are extracted from finite-volume heavy-baryon chiral perturbation theory, $|g_{\Delta N \pi }|=2.80(18)(21)$; similarly for other couplings such as $|g_{\Sigma^* \Lambda \pi }|=2.49(23)(35)$, $|g_{\Xi^* \Xi \pi }|=2.49(23)(35)$.
The advantage of the latter approach is that it is easier to calculate and obtains better signal-to-noise ratio than the direct approach. However, it depends on the reliability and our knowledge of the applicable chiral perturbation theory.
A similar approach might be applied in the near future to find couplings such as $g_{A,N^*}$ and $g_{NN^*\pi}$.

{\bf Large-$Q^2$ Prospects: }%
Significant progress has been made in lattice QCD to achieve large-$Q^2$ form factors. Remaining obstacles are mainly due to the available lattice spacings\footnote{The momenta allowed on the lattice (assuming periodic spatial boundary conditions) are $\frac{2\pi n}{L}a^{-1}$, where $L$ is the length of the spatial volume, while $n$ ranges 0, 1, $\sqrt{2}$, $\sqrt{3}$, etc. Thus, the finer the lattice spacing, the higher momentum transfer one can achieve.} used in the current calculations and
the signal-to-noise ratio at higher momenta for a given lattice spacing.
Refs.~\cite{Lin:2010fv,Lin:2011sa} investigate the latter limitation by reexamining the typical steps in a form-factor calculation: traditional methods optimize lattice parameters to eliminate excited-state contamination for a hadron at rest or low momentum. When one starts to boost the hadron to higher momentum, the hadron energy becomes compatible with the excited-state mass at rest, and the original parameters screen out any useful signal. The solution to this problem is to allow multiple parameter values that overlap with not only the ground-state hadron but also excited ones, and to extend the form-factor analysis to specifically include the excited states at each momentum. Thus, at higher momenta, there would be operators that still have good overlap. Using this approach, the number available form factors can be extended by a factor of 2--3, until the lattice discretization effects at $O((ap)^2)$ become dominant. 
The right-hand side of Fig.~\ref{fig:G1pR-Fpi} displays an example using $N_f=2+1$ pion form factors with 580, 875, 1350~MeV pion masses with $Q^2$ reaching nearly 7~GeV$^2$\cite{Lin:2011sa} for the highest-mass ensemble. The extrapolated form factor at the physical pion mass shows reasonable agreement with JLab precision measurements. Future attempts will focus on decreasing the pion masses and exploring $Q^2$-dependence of pion form factors for yet higher $Q^2$.

Refs.~\cite{Lin:2010fv,Lin:2011sa} also suggest an approach to remove the first obstacle through a lattice technique called ``step scaling'', as demonstrated in Fig.~8 in Ref.~\cite{Lin:2011sa}, using multiple lattice spacings and volumes to reach higher momenta by an order of magnitude.
By calculating the step-scaling function at overlapping momentum points (or interpolating) we can reduce the systematic error due to finite-volume or lattice-discretization artifacts. However, generating several volumes and tuning for multiple lattice spacings of dynamical $N_f=2+1$ lattices requires a large amount of computational resources; we hope such a proposal will become feasible as petascale computing facilities become available in the near future.

{\bf Conclusion and Outlook: }%
There has been much progress in Roper-related quantities calculated from lattice QCD, and the precision and techniques used have been greatly improved over the past couple years. We see impressive excited-baryon spectra from both the HSC and CSSM groups. The goals of the outlook from the previous Nstar Workshop, including the difficult particle identification from irreducible representations of the cubic group, have been implemented and published. Much remains to be done, but the progress already made has been remarkable.

We see that finite-volume effects will play a significant role in the upcoming precision world. We note that even for the nucleon mass, the finite-volume effect is non-negligible for $M_\pi L \approx 4$. With the Roper, the mass shift of the central values could be up to 170~MeV. We encourage the both HSC and CSSM to look for volume corrections in their calculations, either with multiple-volume calculations or estimating them from effective theory.
A volume dependence study could also provide an indirect approach to
extract the axial
coupling constants related to Roper states.

The Roper-nucleon electromagnetic and axial transition form factors also progressed to dynamical lattice calculations. We use large volume anisotropic lattices to avoid finite-volume effects from the Roper state.
The effort has been concentrated on the small-$Q^2$ region, where the sign discrepancy has been observed with respect to the previous quenched studies.
We may have resolved the discrepancy in the small-$Q^2$ region from the CLAS analysis data that we have been seeing in quenched calculations.
A first axial transition form factor is also reported in this proceeding, which will give us insight into spin structure.
Further studies to increase statistics and refine the analysis on the current data are on-going.

{\bf Acknowledgments: }
The authors are supported in part by the U.S. Dept. of Energy under grant No. DE-FG02-97ER4014.
SDC is also in part supported by DOE grant Nos. DE-FG02-91ER40676 and DE-FC02-06ER41440 and NSF grant No. 0749300 during the earlier stage of the work.
The dynamical form-factor data was taken using UW Hyak cluster, NSF MRI PHY-0922770.

\vspace{-0.5cm}

\begin{thebibliography}{14}
\expandafter\ifx\csname natexlab\endcsname\relax\def\natexlab#1{#1}\fi
\providecommand{\enquote}[1]{``#1''}
\expandafter\ifx\csname url\endcsname\relax
  \def\url#1{\texttt{#1}}\fi
\expandafter\ifx\csname urlprefix\endcsname\relax\def\urlprefix{URL }\fi
\providecommand{\eprint}[2][]{\url{#2}}

\bibitem[Mahbub et~al.(2010)]{Mahbub:2010rm}
M.~S. Mahbub, W.~Kamleh, D.~B. Leinweber, P.~J. Moran, and A.~G. Williams
  (2010), \eprint{1011.5724}.

\bibitem[Bulava et~al.(2010)]{Bulava:2010yg}
J.~Bulava, et~al., \emph{Phys. Rev.} \textbf{D82}, 014507 (2010),
  \eprint{1004.5072}.

\bibitem[Edwards et~al.(2011)]{Edwards:2011jj}
R.~G. Edwards, J.~J. Dudek, D.~G. Richards, and S.~J. Wallace  (2011),
  \eprint{1104.5152}.

\bibitem[Beane et~al.(2011)]{Beane:2011pc}
S.~R. Beane, et~al., \emph{Phys. Rev.} \textbf{D84}, 014507 (2011),
  \eprint{1104.4101}.

\bibitem[Lin et~al.(2009)]{Lin:2008pr}
H.-W. Lin, et~al., \emph{Phys. Rev.} \textbf{D79}, 034502 (2009),
  \eprint{0810.3588}.

\bibitem[Lin et~al.(2008{\natexlab{a}})]{Lin:2008qv}
H.-W. Lin, S.~D. Cohen, R.~G. Edwards, and D.~G. Richards, \emph{Phys. Rev.}
  \textbf{D78}, 114508 (2008{\natexlab{a}}), \eprint{0803.3020}.

\bibitem[Lin et~al.(2008{\natexlab{b}})]{Lin:2008gv}
H.-W. Lin, S.~D. Cohen, R.~G. Edwards, K.~Orginos, and D.~G. Richards,
  \emph{PoS} \textbf{LATTICE2008}, 140 (2008{\natexlab{b}}),
  \eprint{0810.5141}.

\bibitem[Mokeev(2011)]{Mokeev:2011}
V.~I. Mokeev, \emph{these proceedings}  (2011).

\bibitem[Mokeev(2010)]{Mokeev:2010va}
V.~I. Mokeev  (2010), \eprint{1010.0712}.

\bibitem[Aznauryan et~al.(2009)]{Aznauryan:2009mx}
I.~G. Aznauryan, et~al., \emph{Phys. Rev.} \textbf{C80}, 055203 (2009),
  \eprint{0909.2349}.

\bibitem[Dugger et~al.(2009)]{Dugger:2009pn}
M.~Dugger, et~al., \emph{Phys. Rev.} \textbf{C79}, 065206 (2009),
  \eprint{0903.1110}.

\bibitem[Mokeev et~al.(2009)]{Mokeev:2008iw}
V.~I. Mokeev, et~al., \emph{Phys. Rev.} \textbf{C80}, 045212 (2009),
  \eprint{0809.4158}.

\bibitem[Lin et~al.(2010)]{Lin:2010fv}
H.-W. Lin, S.~D. Cohen, R.~G. Edwards, K.~Orginos, and D.~G. Richards  (2010),
  \eprint{1005.0799}.

\bibitem[Lin and Cohen(2011)]{Lin:2011sa}
H.-W. Lin, and S.~D. Cohen  (2011), \eprint{1104.4319}.

\end{thebibliography}

\end{document}